\documentclass[letterpaper,twocolumn,10pt]{article}
\usepackage{usenix-2020-09}

\usepackage{tikz}
\usepackage{amsmath}
\usepackage{needspace}
\usepackage[normalem]{ulem}
\usepackage{appendix}
\usepackage{xspace}
\usepackage{fancyhdr}
\usepackage{comment}
\usepackage{enumitem}
\usepackage{multirow}
\usepackage{wrapfig}
\usepackage{balance}
\usepackage{subfig}
\usepackage{duckuments}
\usepackage{xcolor}
\definecolor{aliceblue}{rgb}{0.94, 0.97, 1.0}
\usepackage{tcolorbox}
\tcbset{width=1\columnwidth,boxrule=1pt,colback=aliceblue,arc=1pt,left=2pt,right=2pt,boxsep=0.5pt}

\usepackage[noabbrev,capitalise]{cleveref}
\usepackage[]{hyperref}

\newcommand\revision[1]{{\color{black} {#1}}}
\newcommand\usenix[1]{{\color{black} {#1}}}

\newcommand\TheName{Citadel\xspace}

\newcommand\aegisAvgStranding{{5.8\%}\xspace}
\newcommand\aegisAvgLoss{{1.6\%}\xspace}
\newcommand\aegisAvgOverhead{{7.4\%}\xspace}
\newcommand\silozAvgOverhead{{33\%}\xspace}

\newcommand*\circled[1]{\tikz[baseline=(char.base)]{
            \node[shape=circle,draw,inner sep=0.5pt, fill=white, text=black] (char) {#1};}}

\Crefname{figure}{Fig.}{Figs.}
\crefname{figure}{Fig.}{Figs.}
\crefformat{section}{#2§#1#3}
\crefformat{subsection}{#2\S#1#3}
\crefformat{subsubsection}{#2\S#1#3}

\begin{document}

\date{}

\title{\Large 
Preventing Rowhammer Exploits via Low-Cost Domain-Aware Memory Allocation
}

\author{
{\rm}
\and
{\rm}
\and
{\rm Anish Saxena}
\and
{\rm Walter Wang}
\and
{\rm Alexandros Daglis}
\and
{\rm}
\and
{\rm}
\and
{\rm \textit{Georgia Institute of Technology}}
} %
\maketitle

\begin{abstract}

Rowhammer is a hardware security vulnerability at the heart of every system with modern DRAM-based memory. Despite its discovery a decade ago, comprehensive defenses remain elusive, while the probability of successful attacks grows with DRAM density. Hardware-based defenses have been ineffective, due to \revision{considerable cost,} delays in commercial adoption, and attackers' repeated ability to circumvent them. %
Meanwhile, more flexible software-based solutions either incur substantial performance and memory capacity overheads, or offer limited forms of protection.

\TheName is a new memory allocator design that %
\revision{prevents Rowhammer-initiated security exploits}
by addressing the vulnerability's root cause: physical adjacency of DRAM rows. \TheName enables creation of flexible security domains and isolates different domains in physically disjoint memory regions, \textit{guaranteeing security by design.} On a server system, \TheName supports thousands of security domains at a modest \aegisAvgOverhead average memory overhead and no performance loss.  In contrast, recent domain isolation schemes fail to support many workload scenarios due to excessive overheads, and incur 4--6$\times$ higher overheads for supported scenarios. 
\revision{As a software solution, \TheName offers readily deployable Rowhammer-aware isolation on legacy, current, and future systems.}

\end{abstract}

\section{Introduction}
\label{sec:intro}

DRAM technology has been scaling to support the growing capacity needs of modern applications. As process nodes get smaller, DRAM bit cells are packed more closely for higher density. The resulting increase in inter-cell interference exacerbates Rowhammer~\cite{kim2014flipping}, in which frequent DRAM row activations cause bit flips in nearby rows. Rowhammer is a severe security threat as bit flips in critical data can compromise the entire system~\cite{seaborn2015exploiting} and exploits can be launched across VMs~\cite{xiao2016bitflipflops}, from unprivileged software~\cite{van2016drammer}, remote code over the network~\cite{nethammer}, and malicious websites~\cite{gruss2016rhjs}.

Rowhammer affects DRAM chips spanning all major technology nodes~\cite{kim2020revisitingRH}. 
A 2022 study found \textit{all} tested DDR4 chips vulnerable~\cite{jattke2021blacksmith}, despite presence of Rowhammer protection in hardware~\cite{frigo2020trrespass}. %
More principled hardware mitigations have been proposed~\cite{saxena2022aqua, yauglikcci2021blockhammer, kim2021mithril}, but are currently not  deployed~\cite{loughlin2023siloz}. 
In contrast, software-based mitigations can adapt to tackle the evolving landscape of Rowhammer exploits without any hardware changes. Such solutions typically prevent \textit{exploitation} of bit flips, like detecting bit flips in page tables to prevent privilege escalation~\cite{SoftTRR}, demoting Rowhammer from a security threat to a reliability concern. However, such specialized schemes do not generalize to protect arbitrary data.

\begin{figure*}
    \centering
     \includegraphics[width=0.95\textwidth]{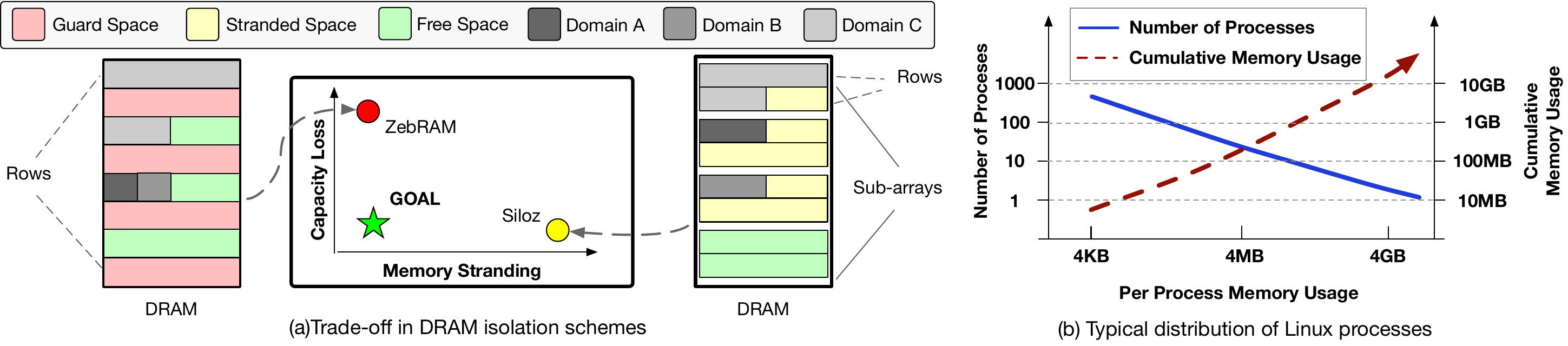}    
     \caption{(a) Row-sized chunks (left) waste memory capacity (67\% loss with two guard rows); subarray-sized chunks (right) strand memory (\silozAvgOverhead in our evaluations) and limit the number of supported security domains. (b) In typical systems, the memory footprint of most processes is below a few megabytes, while most memory is occupied by few large-footprint processes. 
     } 
    \label{fig:intro}
\end{figure*}

A category of software solutions opts for a principled domain isolation approach~\cite{loughlin2023siloz, van2018guardion, bock2019rip}, where untrusting domains are physically isolated from each other in memory.
Breaking the physical proximity in DRAM prevents aggressors in one domain from impacting other domains. 
For example, each DRAM row containing data can be flanked by guard rows (which are inaccessible by applications) on both sides, to absorb any Rowhammer-induced bit flips  (\cref{fig:intro} (a)-left).
ZebRAM~\cite{konoth2018zebram} leverages this mechanism to stripe the memory into alternating data rows and guard rows, %
but incurs an impractical 50\% DRAM capacity loss, or more, to defend against complex attack patterns~\cite{half-double}). %

Another recent solution, Siloz~\cite{loughlin2023siloz}, avoids memory capacity loss by foregoing guard rows for most domains.
Instead, Siloz reserves entire DRAM sub-arrays---which comprise hundreds of rows each and are physically isolated---to individual domains (\cref{fig:intro} (a)-right). 
While coarse-grained gigabyte-scale domains are suitable for Siloz' intended use-case of VM isolation, the approach limits the maximum supported number of domains and fails to generalize: domains with small memory footprint underutilize their reserved memory that cannot be allotted to other domains, leaving it \textit{stranded}~\cite{pond_micro_stranding}.
\textit{Memory stranding} can reduce system utilization, because memory reserved for one domain cannot be used to satisfy memory requirements of another domain.
In evaluations with SPEC-2017~\cite{SPEC2017} and Graph Analytics~\cite{GAP} workloads on a server system, Siloz strands \silozAvgOverhead of the 128GB memory on average, which prevents it from supporting several workload scenarios even when requisite domains are available. %
\newpage
\textit{Overall, current isolation schemes are unsuitable for widespread adoption due to overly broad domain definition (like entire VMs) and unacceptable memory capacity overheads in two forms: stranding, and loss due to guard rows.}

To ensure future-proof protection against both known and yet unknown Rowhammer exploits, an isolation-based scheme should ideally support any number of software-defined domains of arbitrary size.
Such software-defined domains must incur low runtime and memory overheads for practical adoption.
We enable such a solution with \TheName, a new memory allocator
which supports thousands of variably sized domains at no performance cost and modest memory overheads.
Unlike current solutions that either isolate KB-sized pages (like ZebRAM) or coarse-grained GB-sized regions (like Siloz), \TheName is well-suited to---but is not limited to---the natural isolation granularity of individual processes.

Supporting requisite domains requires sizing the memory reservation primitive properly, to balance the delicate tradeoff between memory loss and stranding. 
Increasing core counts (e.g., 512 hyperthreads on a 2-socket AMD Bergamo processor) and Rowhammer exploits on sub-process targets~\cite{gruss2016rhjs, nethammer} necessitate provisioning several hundreds or thousands of security domains.
In addition to hundreds of such \textit{large} domains, many Linux processes in typical systems (kernel helpers and daemons) use \textit{few MBs of memory or less}, as shown in \cref{fig:intro} (b).
Such processes can be targeted~\cite{throwhammer} and require thousands of \textit{MB-sized} domains for protection.
Preventing attacks on page tables~\cite{zhang2020pthammer} also requires a separate domain for each page table page, resulting in tens of thousands or more of \textit{KB-sized} domains.
Hence, our design needs to incur low memory overheads while supporting \textit{both} large workload domains and numerous smaller-footprint bespoke domains.

We achieve this dual goal by leveraging two key observations: (i) domains co-located within the same row and flanked by guard rows cannot hammer each other, and (ii) the cumulative memory footprint of small domains is small compared to total memory requirement (\cref{fig:intro} (b)).
\TheName, therefore, employs a two-level design where allocation requests are first allotted from a small, dynamically sized region of striped memory, which incurs high relative capacity loss but supports millions of small domains.
Once a domain exceeds a tunable memory footprint threshold, subsequent allocations are handled by the next level, which uses \textit{reservation chunks}.
A reservation chunk consists of a configurable number of contiguous rows and is sized to amortize the capacity loss due to guard rows.
Our chosen chunk size of 16 rows bounds worst-case capacity loss at 12.5\% (with two guard rows) while still supporting thousands of domains. 

\revision{
We develop the \TheName memory allocator and integrate it to the Linux kernel (v5.15). 
We evaluate with dynamic mixes of SPEC~\cite{SPEC2017} and GAP~\cite{GAP} workloads on a server system running Ubuntu 20.04.
\TheName provides Rowhammer resilience with only \aegisAvgLoss capacity loss and \aegisAvgStranding stranding on average (7.4\% total memory overheads), 4--6$\times$ lower than prior works. Even with workloads using {85\%} of memory, the total overhead remains low at {5\%}, making \TheName a low-cost and secure solution.
We also provide guidance on tackling important deployment challenges using our design.
For example, our design elegantly maps software domains to the hardware layout of memory.
Our principled approach is applicable to various complex DRAM mappings prevalent in modern servers.
Moreover, while memory sharing is typically disallowed to avoid inter-domain exploits (and disabled in our prototype), we describe how \TheName's striped memory region can be leveraged to allow for safe memory sharing at low cost.
}

In summary, our paper makes the following contributions:
\begin{enumerate}[noitemsep,topsep=0pt,leftmargin=*]
    \item We design a memory allocator that provides guaranteed Rowhammer resilience at low overheads, irrespective of the number or size of security domains.
    \item We balance guard row capacity loss and stranding overheads by leveraging the natural granularity of DRAM isolation with the \textit{reservation chunk} primitive. 
    \item We develop a two-level design that offers complete protection coverage against future unknown attacks by supporting thousands (and even millions) of security domains.    
    \item We provide an implementation of \TheName in the Linux kernel that is secure and incurs no performance overhead. 
\end{enumerate}

\section{Background}
\label{sec:background}

\subsection{DRAM Organization}

DRAM-based memory is structured hierarchically into channels, sub-channels, banks, and rows. DDR5 organizes each 64-bit channel into two independent 32-bit sub-channels, each with a burst length of 16 to deliver a 64B cache line. A sub-channel \revision{controls up to} 32 banks arranged in an array of rows and columns, with a typical configuration being 128K rows of 8KB each. 
While the DDR interface necessitates usage of banks and ranks, internally, the DRAM microarchitecture further distributes rows in each bank into electrically independent sub-arrays, each comprising 512--2K rows (i.e., 64--256 sub-arrays per bank).
\revision{Moreover, each 8KB DRAM row is split into two half-rows, and both half-rows are simultaneously utilized.}
Retrieving data from DRAM requires the relevant row in a bank to be \textit{activated}, bringing the data into the row buffer. Accessing data in a different row involves clearing the row buffer with a ``precharge'' command, followed by the activation of the newly requested row.

\subsection{The Rowhammer Vulnerability}

Rowhammer is a data-disturbance bug wherein frequent DRAM row activations cause bit flips in nearby rows. Since Rowhammer was first characterized in 2014, it has been demonstrated in memory from all major vendors, spanning all major technology nodes, and across thousands of DRAM chips~\cite{kim2020revisitingRH}. In fact, a recent study from 2022~\cite{jattke2021blacksmith} found that \textit{all} recent DDR4 chips tested are vulnerable.

Rowhammer is not just a reliability concern, but a severe security threat. Rowhammer can be launched from across VMs~\cite{xiao2016bitflipflops}, unprivileged software~\cite{van2016drammer}, remote code over the network~\cite{nethammer}, and even malicious websites~\cite{gruss2016rhjs}, causing significant damage. Flipping bits in page tables leads to privilege escalation attacks~\cite{seaborn2015exploiting}. Data-dependence of bit-flips can be exploited to leak sensitive data from victim rows located near an adversary~\cite{kwong2020rambleed}. Highly repeatable and targeted bit flips bypass authentication mechanisms~\cite{gruss2018another} and break cryptosystems~\cite{flipfengshui}. There are numerous other exploits~\cite{hammerscope, glitch_rh,  sgxbomb,  throwhammer}, making Rowhammer protection in software a moving target.

\subsection{Threat Model}
\label{sec:threat-model}

We assume the system has a typical Operating System (OS) with virtual memory and page tables providing process isolation. 
The system uses DRAM main memory which is vulnerable to Rowhammer. 
We assume the kernel code is trusted and runs correctly. 
The attacker runs code with \textit{user} privileges (either natively or over the network, via websites, inside VMs, etc.) and leverages Rowhammer to flip bits to damage data integrity, leak information, break confidentiality guarantees, or achieve privilege escalation.
To prevent exploitation of Rowhammer, we assume the OS employs isolation of security domains in physical memory to disallow a domain from flipping bits in another domain.
Without loss of generality, we assume each process is untrusting of each other, thus at least one security domain per process is required.
Moreover, each page table page also requires its own security domain to prevent implicit attacks~\cite{zhang2020pthammer}.
Finally, given that exploits can arise from even within the same process (e.g., from malicious websites~\cite{gruss2016rhjs}), we want to protect arbitrarily defined domains. 
An exploit is successful if an attacker can flip bits in any domain not belonging to the attacker's domain.

\subsection{Memory Mapping}
\label{sec:mapping}

\begin{figure}
    \centering    
           \includegraphics[width=.7\columnwidth]{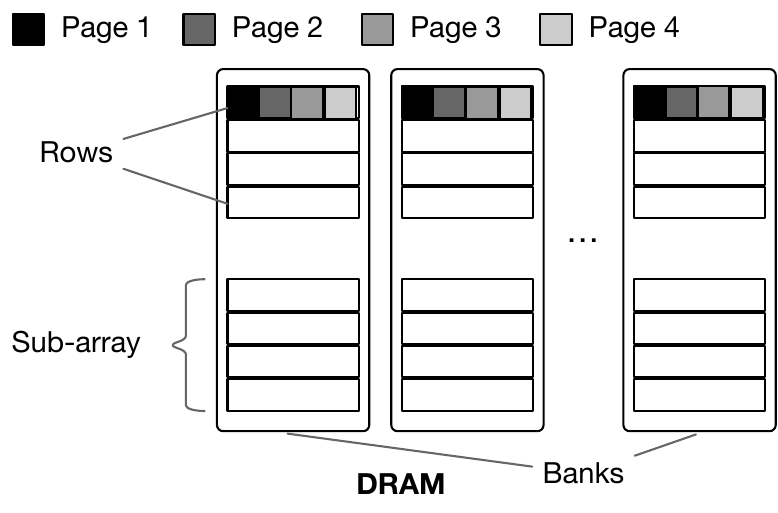}
  \vspace{-2mm}
    \caption{Modern memory mappings distribute lines belonging to a page to different banks of memory. 
    }
    \label{fig:mem_mapping}
\end{figure}

The OS views physical memory as a linear address range which it subdivides into page frames, typically of size 4KB.
In the hardware, 64B-lines belonging to the frame (64 lines for a 4KB page) are mapped to a DRAM channel, rank, bank, row, and column by a mapping function.
Lines of a page in a bank are placed in same DRAM row, aiding row buffer hits.
Spatially proximate lines from contiguous pages are distributed to different banks, hiding activation latency via bank-level parallelism within the channel~\cite{mop}.
Moreover, 
bank selection is derived via XOR-based address hashing.

\cref{fig:mem_mapping} shows the effect of common mappings on the layout of OS-visible pages to DRAM.
Even if processor vendors are unwilling to disclose the memory mapping function, they can be reverse-engineered, as often done by attackers before launching Rowhammer attacks~\cite{jattke2021blacksmith, wang2020dramdig}.
Note that while an individual 4KB page spans multiple banks, it does \textit{not} cover multiple rows within the bank, and the set of rows across banks comprising portions of the same OS page share the same row ID.
Therefore, given knowledge of the memory mapping function, we can derive the row ID that represents the physical placement of each page frame in memory.
As found in prior work \cite[\S4.2]{loughlin2023siloz}, 2MB pages may exceed a single global row, but will map to consecutive global rows, while this is not true for 1GB pages.

The DRAM might internally remap rows in a bank, which complicates, but does not prevent, the deduction of physical to device address mapping, and thus knowledge of the location of page frames in DRAM.
As done in prior work, %
internal remapping of faulty rows can be identified and removed from allocatable memory to preserve isolation~\cite{loughlin2023siloz}.
\revision{We discuss implications of complex DRAM addressing in \cref{sec:complex_addressing}.}

\subsection{Memory Allocation in Linux}

The OS 
allocates page frames to all processes.
Linux uses the buddy allocator that allocates physical memory in powers of 2 of the base page size, which is known as the order of allocation (a single 4KB page is order 0). 
The maximum order is 11, in which case buddy allocates a contiguous memory region of size $2^{11 - 1}\times 4KB$ or 4MB.
The memory subsystem in Linux determines the order of pages to be allocated on an allocation request.
The buddy allocator maintains free lists for each allocation order, comprising physically contiguous memory regions of that size (4KB, 8KB, and so on).
As higher-order lists are depleted over time, memory becomes fragmented, and the memory subsystem resorts to order-0 requests for most allocations.
Recent work finds that memory in servers gets fragmented within one hour of running workloads~\cite{zhao2023contiguitas}; we use this observation to ensure our experiments are representative of real-world usage by running dynamically spawning workload mixes for several hours.

\section{Software-based Rowhammer Mitigations}
\label{sec:software-defense}

Despite a decade of research and mitigation attempts by the industry~\cite{frigo2020trrespass, jattke2021blacksmith, half-double}, Rowhammer persists in modern systems.
JEDEC, the organization regulating the DDR standard, states that ``in-DRAM mitigations cannot eliminate all forms of Rowhammer attacks''~\cite{JEDEC-RH1,JEDEC-RH2}. 
Principled hardware solutions with low runtime overheads~\cite{park_graphene:_2020, qureshi2022hydra} remain elusive due to implementation costs, while low-cost approaches~\cite{samsung_dsac, isscc23} can be circumvented by complex attack patterns.
In contrast, software-based solutions are attractive as they can adapt to the rapidly evolving landscape of Rowhammer attacks.

The OS has limited view into memory accesses, so while software-based Rowhammer detection schemes exist~\cite{aweke2016anvil}, they %
either provide insufficient protection or do not generalize to protect all data.
System-wide software solutions, instead, can opt for domain isolation~\cite{bock2019rip, brasser2017can, van2018guardion}, where untrusting domains are isolated from each other in memory, preventing one domain from impacting other domains. Such schemes make memory allocation DRAM-and-Rowhammer-aware, demoting Rowhammer from a security threat to a reliability concern.
\revision{Notably, while Rowhammer persists despite commercial mitigations~\cite{frigo2020trrespass, samsung_dsac}, the underlying adversarial patterns are absent in benign workloads. 
By eliminating Rowhammer's security threat, such solutions relegate reliability concerns to DRAM, where they are best addressed.
}

\subsection{ZebRAM: Guard-row Based Protection}
\label{sec:zebram}

ZebRAM~\cite{konoth2018zebram} isolates every page by striping the memory into alternating guard and data rows.
As pages co-located within the same data row cannot hammer each other, and guard rows are not accessible from userspace, bit flips cannot occur in the data region.
Untrusting applications need not be explicitly assigned to any domain, since all data rows are immune to bit flips.
ZebRAM was tested with a single guard row per data row, resulting in 50\% capacity loss. 
To avoid wasting half of the memory, ZebRAM uses the guard rows as swap space with either error correction or integrity protection to protect ``swapped-out'' data.
As a result, half of the memory becomes slow, integrity-protected swap space.
When the resident working set exceeds 50\% of the memory, ZebRAM incurs impractical slowdown---e.g., $3\times$ slowdown at 70\% memory usage. 
Thus, in practice, only half of the DRAM capacity is available when performance matters. %

\subsection{Siloz: Sub-array Based Protection}

Siloz~\cite{loughlin2023siloz}, avoids the overhead of guard rows by using sub-array-based isolation.
Because DRAM sub-arrays are physically independent structures, Rowhammer interference across sub-arrays is impossible.
Siloz leverages this to assign entire DRAM sub-arrays to a single security domain.
All rows in a sub-array are \textit{reserved} for the same domain, preventing bit flips from propagating across domains, ensuring isolation.
As conventional mapping spread lines of pages across banks, Siloz collates the group of sub-arrays from all banks in memory sharing the same sub-array index.
With 128K-row banks, a sub-array of 1K rows enables 128 sub-array groups, allowing for 128 distinct, equally-sized domains (each allotted  $\sim\frac{1}{128}{th}$ of the memory capacity).
For Siloz' target of VMs with multi-GB footprints, GB-sized sub-arrays provide reasonable memory reservation granularity.

\subsection{Limitations of Domain-isolation Solutions}
\label{sec:prior-limitations}

While domain isolation provides robust security guarantees, current solutions are either tailor-made for specific use cases 
or employ drastic measures to provide protection 
, impeding their practical or widespread adoption.

\smallskip
\noindent \textbf{Drastic memory capacity loss.} ZebRAM, with its striped memory pattern, effectively reduces memory capacity by 50\%. 
Exploits like Half-Double~\cite{half-double} have shown bit flips in neighbors two rows away from the aggressor, necessitating two guard rows for 67\% capacity loss.
Mitigating this loss by using guard rows as swap space is no longer viable as the Rowhammer threshold shrinks.
First, as up to seven bit-flips within 8-Byte word have been observed~\cite{de2021smash}, data in swap space can be corrupted even when integrity failures are detected.
Second, it is easier to hammer the swap space itself to cause bit flips in data rows, because reading out a page from the swap space is a $\mu$s-level operation, and just a few thousand activations are enough to cause bit flips~\cite{kim2020revisitingRH} in modern memory.
Thus, guard rows cannot hold data, making ZebRAM-like solutions impractical for system-wide use.

\smallskip
\noindent \textbf{Insufficient number of domains.} Siloz minimizes guard row loss by leveraging the natural isolation provided by DRAM sub-arrays. However, the limited number of sub-arrays per bank restricts the number of supported domains. 
Unfortunately, Rowhammer has been shown to break many software-defined boundaries, including VM isolation, kernel-userspace isolation, process isolation, and even sandboxing for websites \cite{gruss2016rhjs}, thus requiring the creation of increasingly numerous domains to protect against new exploits.
To support such scenarios, Siloz would need to coalesce untrusting entities into larger domains, jeopardizing system security. 

\smallskip
\noindent \textbf{Excessive memory stranding.} While Siloz avoids memory loss to guard rows, it still incurs severe memory overheads.
For domains that do not use the entirety of a GB-sized sub-array group, the reserved memory becomes effectively stranded.
Stranded memory is only available to a domain, but unusable by other domains that may require it.
In our experiments with composite SPEC and graph workloads on a 48-core server system, Siloz cannot support some scenarios even when requisite domains and memory capacity are available, due to excessive stranding, which is \silozAvgOverhead of the 128GB memory on average (computed using supported scenarios). 

\smallskip
\noindent \textbf{Other considerations.} An alternative approach of isolating different domains in individual memory banks is incompatible with modern memory mappings, which stripe each page across many banks.
Further, it would compromise single-domain performance by foregoing bank-level parallelism.

\subsection{Goal of Our Paper}

We target a domain isolation solution that supports thousands of arbitrarily sized domains and is compatible with existing hardware, while minimizing memory capacity loss, memory stranding, and performance loss. We develop such a solution with our new memory allocator, \TheName.

\section{Efficient Domain Isolation with \TheName}
\label{sec:design}

Domain isolation solutions do not scale to support arbitrary number and size of domains due to a combination of high memory overheads and limited number of supported domains. 
The key to minimizing memory overheads---loss and stranding---is to reserve enough memory to a domain (even if not all of it is initially allocated) to amortize the cost of guard rows, while minimizing stranding of reserved memory due to the dynamic behavior of memory allocation/deallocation within the reserved region.
In this section, we describe our new memory allocator, \TheName, that achieves this goal by choosing the correct \textit{memory reservation primitive}, which also dictates the number of supported domains.

\subsection{Software-Defined Security Domains}

A \textit{security domain} comprises a memory footprint that must be protected from Rowhammer attacks by entities outside the said domain.
Examples of such domain include a process, a thread of a process, a page table page, a sandboxed region in a browser, etc.
Memory allocation can involve interleaved requests from many domains at runtime, so a given domain may occupy non-contiguous regions of physical memory. Each such region forms a \textit{security zone}.
A {secure} memory allocation involves allocating memory from a zone that has previously been \textit{reserved} for the requester's domain.
In \TheName, every page allocation request is associated with a domain-ID and memory is provisioned only from the domain's security zones (details in \cref{sec:impl}), ensuring secure memory allocation.

\subsection{Mapping Domains and Zones to Memory}

We define a \textit{global row} as the set of DRAM rows sharing the same row ID across all banks in \revision{a NUMA domain's} memory. 
Global row is the semantic bridge between the OS-centric and hard\-ware-centric view of memory, to account for the physical distribution of a single logical page to multiple different banks in memory.
We control the physical placement and neighborhood of an OS page in memory by using global row as a basic unit of memory reservation.
For example, if a security zone reserves global row $X$, then subsequent allocation requests can be satisfied by allotting pages that map to global row $X$. Moreover, we can place \textit{guard rows} that are not mapped to any domain, at global rows $X\pm1$, to ensure isolation against all other domains in memory. 

In \TheName, each security zone is a contiguous range of global rows, starting with one or more guard rows followed by global rows where application data can be allocated.
Guard rows, that prevent propagation of Rowhammer across zones, are not mapped to any domain (conversely, no page maps to guard rows) and are never accessible by an adversary.

As discussed in \cref{sec:mapping}, each 4KB page is distributed within a single global row, while 2MB pages span consecutive global rows, thus both can be naturally contained within a zone.
Because 1GB pages require finding gigabyte(s) of contiguous memory and complicates zone definition, our current design does not support it for simplicity. We note that finding large contiguous memory regions would be more challenging in the baseline~\cite{zhao2023contiguitas} than in \TheName, whose coarser-grained memory reservation reduces fragmentation.

\subsection{Memory Chunk Reservation Primitive}

A configurable number of contiguous global rows comprises a \textit{reservation chunk}, \TheName's unit of memory reservation.
Memory reservation is effectively a pre-allocation of a memory region that can be used by a given domain while being unavailable to any other domain.
Each zone may comprise \revision{multiple} contiguous chunks, \revision{and only the first chunk of a given zone needs guard rows.
Hence, the chunk size (i.e., the number of global rows in a chunk) and the resulting number of chunks per zone impact the memory overheads, as we elaborate in \cref{sec:chunk-size-tradeoffs} and \cref{sec:design:zone_exp}, respectively.}

\begin{figure}
    \centering
    \includegraphics[width=0.45\textwidth]{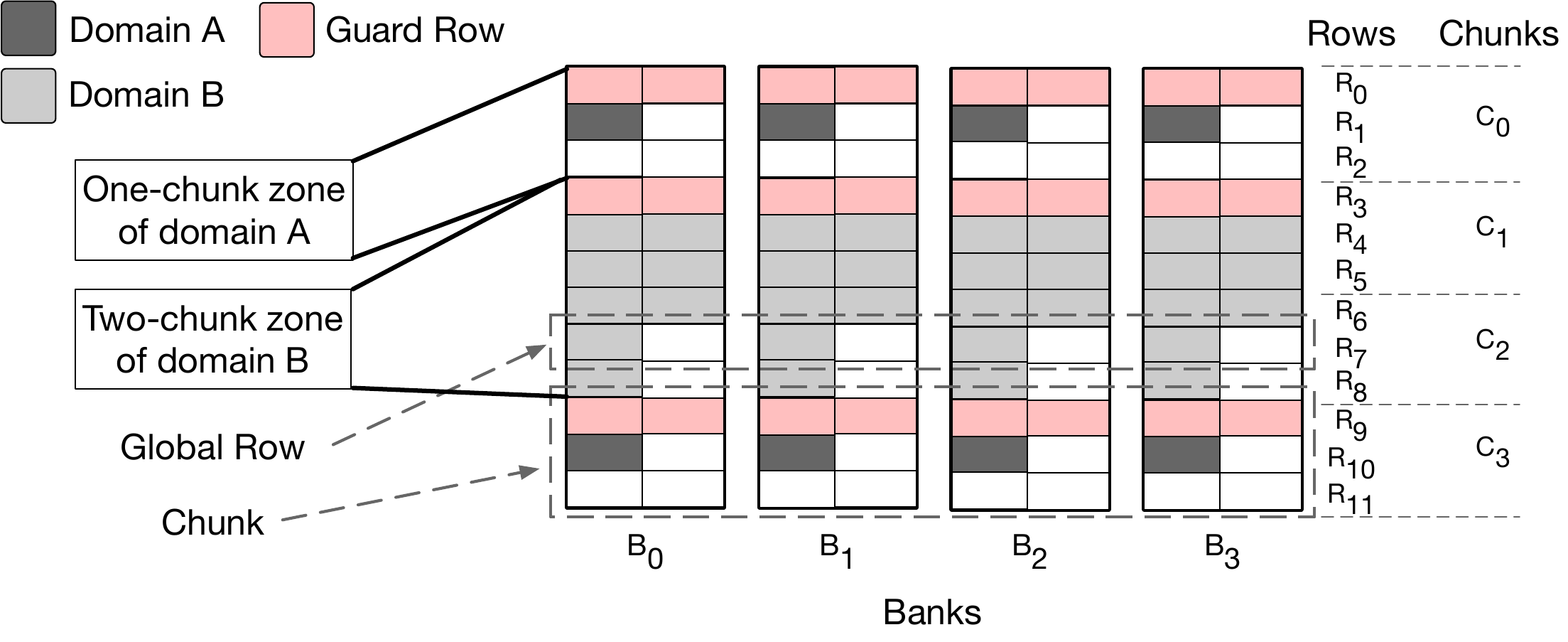}
     \vspace{-2mm}
    \caption{Illustration of security domains and corresponding security zones in memory. Each zone comprises a contiguous range of global rows, with the first $N_G$ of them serving as guard rows. One guard row per zone ($N_G=1$) shown.%
    }
    \label{fig:terminology}
\end{figure}

\cref{fig:terminology} depicts two security domains, A and B, assuming a chunk size of three global rows.
Domain A consists of two single-chunk-sized zones, while domain B consists of a single zone of two contiguous chunks.
Each \textit{zone} features a guard row; \revision{zone B's two contiguous chunks do not require guard row separation}.
Global row $R_1$ stores one OS page of domain A that physically straddles banks $B_0$ to $B_3$. 
Similarly, domain B's zone contains eight OS pages spread across four banks and five rows, and can accommodate two more pages if needed. 
\revision{Domain B's contiguity results in higher efficiency. Of the six global rows each domain occupies, A's two needed guard rows leave it with four usable rows, one less than B.}

\subsection{Chunk Size and Memory Overheads}
\label{sec:chunk-size-tradeoffs}

The reservation chunk is the minimum granularity of memory reservation for a given domain (disregarding zonelets, described in \cref{sec:design:zonelets}).
The chunk size determines the number of supported domains and memory overheads.

\medskip
\noindent\textbf{Impact on Number of Supported Domains.}
Each chunk can only be reserved for one domain, so the maximal number of supported domains is the number of chunks in memory.
Thus with 128K global rows, at most $\frac{128K}{chunk\ size}$ domains are supported.
For example, Siloz' sub-array approach effectively results in a chunk size of 512 global rows (size of the sub-array), and therefore supports a maximum of 256 domains.
A smaller chunk size supports more domains, which we want, but incurs higher capacity loss due to guard rows.

\medskip
\noindent\textbf{Impact on Guard Row Capacity Loss.}
The number of guard rows $N_G$ associated with each chunk depends on DRAM's vulnerability to Rowhammer.
Due to Half-Double~\cite{HalfDouble}, at least two guard rows must be provisioned per zone.
Amortizing cost of guard rows requires larger chunk sizes; for example, a chunk size of 32 with $N_R=2$ has a maximum capacity loss of 6\%.
However, a very large chunk size (say, 512 rows used in Siloz) increases  memory stranding.

\medskip
\noindent\textbf{Impact on Memory Stranding.}
The OS allocates memory using 4KB pages (with order-zero allocations, which are most common), while memory reserved for a domain is in multiples of chunk-size.
A chunk can only be freed if all pages in it have previously been deallocated.
Therefore, a domain's reserved memory typically exceeds its allocated memory.
Reserved memory that has not been allocated is considered stranded memory, as it restricts memory usability to the chunk's associated domain. %
Naturally, the larger the chunk, the higher the potential for memory stranding.
Thus, the choice of chunk size introduces a tradeoff between the two types of memory capacity overheads: memory loss and memory stranding.

\medskip
\noindent\textbf{Choice of Optimal Chunk Size.}
With 128K global rows in memory, supporting more than a thousand domains requires a chunk size smaller than 128.
With two guard rows per chunk, small chunk sizes would incur drastic worst-case guard row capacity loss (e.g., 50\% with 4 rows).
Thus, middling chunk sizes of 8, 16, 32, or 64 rows are suitable candidates to enable efficient domain isolation. 
While we defer the detailed analysis to \cref{sec:eval}, we find that chunk size of 16 global rows strikes the best balance between memory capacity loss and stranding.

\subsection{Reducing Guard Row Loss}
\label{sec:design:zone_exp}

In \TheName, if two adjacent chunks belong to the same domain, then there is no need for guard rows to isolate data rows in these chunks, because the security guarantee of domain isolation protects domains from each other, not from bit flips within the same domain.
Therefore,  \TheName attempts to \textit{expand} one of the pre-existing zones of a growing domain before creating a new zone, to minimize capacity loss.
By reserving an available chunk that is physically adjacent to an existing zone, larger zones can be formed.
With guard rows only required at the start of the zone's first chunk, zone expansion amortizes the cost of $N_G$ guard rows over multiple chunks, reducing the capacity loss on guard rows.

\begin{figure}
    \centering    
           \includegraphics[width=.95\columnwidth]{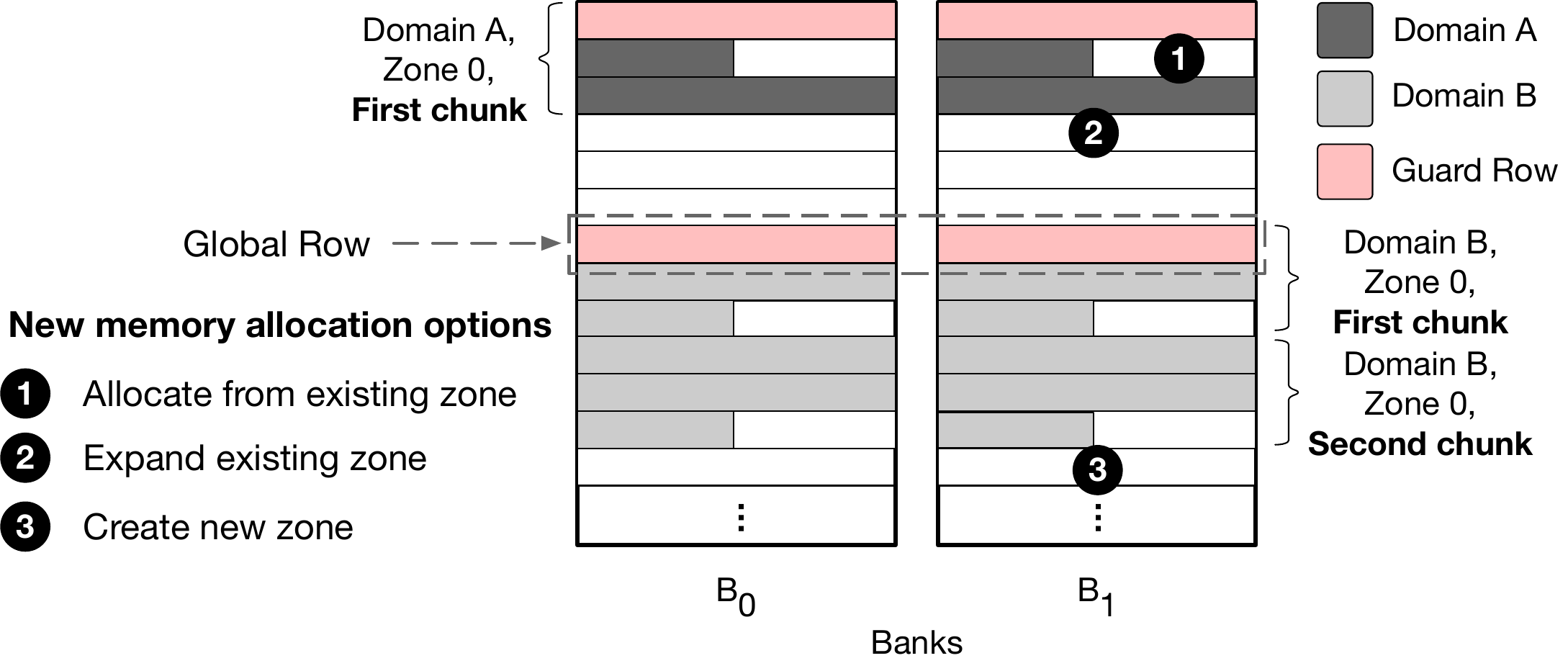}
		 \vspace{-2mm}    
    \caption{Memory allocation and chunk reservation optimizations to minimize memory capacity overheads. $N_G=1$ and two DRAM banks shown for simplicity.}
    \label{fig:optimizations}
\end{figure}

In \cref{fig:optimizations}'s example, upon arrival of new memory allocation requests for domain A, \TheName will first fill the unallocated space available in domain A's already allocated zone 0 (\circled{1}). If that space is not enough, \TheName will expand domain A's zone 0, if there is adjacent space for an additional chunk available (\circled{2}). \cref{fig:optimizations} exemplifies such a successful instance for domain B's two-chunk zone 0; note that only the first chunk requires guard row(s). Finally, if none of the two options can be used to satisfy domain A's memory allocation request, a new zone for domain A will be created (\circled{3}).

\subsection{Zonelets to Protect Millions of Domains} 
\label{sec:design:zonelets}

Zones amortize memory capacity loss for typical domains like mainstream applications with resident memory footprint of tens to hundreds of megabytes or more. 
However, process-level domains are insufficient because Rowhammer can occur even within the same process~\cite{gruss2016rhjs}.
To protect against such exploits, vulnerable memory regions must be designated as a bespoke domain by the programmer.
However, a vulnerable region smaller than a chunk causes memory stranding.
For example, if only one global row within the chunk is allocated, 93\% of the memory in a 16-row chunk will be stranded (with $N_G=2$).  
The need for many small bespoke domains is common, for two reasons. First, they are essential for isolating each page table page, to protect against implicit attacks like PTHammer~\cite{zhang2020pthammer}. Second, a plethora of background processes, which can often be in the hundreds, usually have a memory footprint of a few MBs or less, as shown in \cref{fig:intro} (b). 

\TheName, therefore, supports the additional \textit{zonelet} primitive, which consists of one global data row, along with $N_G$ guard rows. Zonelets incur a high capacity loss (e.g., 66\% with $N_G=2$), but allow safe data colocation of multiple domains within the same row. 
\TheName dynamically provisions zonelets by choosing an available chunk and striping it into alternating ($N_G$) guard rows and data rows. 
Thus, every chunk can be used as $\lfloor\frac{chunk\_size}{N_G+1}\rfloor$ zonelets, practically eliminating memory stranding with controlled capacity loss. 
\revision{In the limiting case, if only zonelets are provisioned, \TheName can support 10.48 million 4KB domains, although at the worst-case capacity loss of 68.8\%, similar to ZebRAM.}
The security guarantees of zonelets and zones are equivalent, \revision{with respect to Rowhammer,} and we leverage these two constructs to reduce Rowhammer from a security threat to a reliability concern.

\subsection{Security Analysis of \TheName}
\label{sec:sec_analysis}

The security guarantee of \TheName is that no Rowhammer-induced bit flips may occur in a given domain due to frequent row activations in any other domain.
\TheName achieves this by physically isolating different domains in distinct regions of memory, based on the following \textbf{principle assumption}:

\begin{tcolorbox}[boxrule=1pt,left=5pt,right=5pt,top=1.5pt,bottom=1.5pt]
\textit{Rowhammer effects are experienced by rows at a distance of at least one and at most $N_G$ away from an activated row.}
\end{tcolorbox}

\TheName provides Rowhammer immunity between domains by ensuring that data belonging to different domains are separated by at least $N_G$ rows, as we show next.

\subsubsection{Security Guarantees of Zones}

We show \TheName is secure by proving the following lemma:

\begin{tcolorbox}[boxrule=1pt,left=5pt,right=5pt,top=1.5pt,bottom=1.5pt]
\textbf{Lemma-1: }%
\revision{\textit{\TheName isolates domains either by colocating them in the same global row (where Rowhammer can't occur, since row distance=0), or by separating them by at-least $N_G$ guard rows which are not mapped to any domain.}}
\end{tcolorbox}

\TheName satisfies each page allocation request with memory belonging to a security zone associated with the requester's domain.
The first $N_G$ global rows in a zone are devoid of data and are not mapped to any process (i.e., no page table entry points to pages in these rows).
Thus, two zones in memory are always separated by $N_G+free\_chunks\times N_G$ unmapped global rows (as each chunk's min size is $N_G$ rows), where $free\_chunks$ are free chunks that may exist between zones.
Even if $free\_chunks=0$, data of different zones is always separated by at least $N_G$ guard rows, proving Lemma-1.
From Lemma-1 and our assumption, it follows that no domain can impact any other domain with Rowhammer.

\subsubsection{Security Guarantees of Zonelets}

Each zonelet is striped into alternating ($N_G$) guard rows and one data row.
Thus, each global data row is separated by $N_G$ guard rows, providing inter-domain Rowhammer immunity due to our assumption.
Note that domains co-located within the same global row cannot hammer each other, so the optimization does not impact the security of \TheName.

\subsection{Mapping Software to Security Domains}
\label{sec:design:primitives}

\TheName requires associating each memory allocation request with a specific security domain. 
We can associate each process or group of jointly trusted processes with a different security domain. The kernel space of each process, including page tables, must also be isolated in a separate security domain. Moreover, the memory allocator can provide an API \revision{(downstreamed via \texttt{malloc}, %
\texttt{mmap}, and/or \texttt{madvise}, for example)} allowing applications to explicitly request placement of specific data structures in a separate domain from the rest of the process.
To allow a process to own multiple domains and to explicitly indicate the domain to be used for each memory allocation request, the kernel must perform additional domain use permission checks before serving the request.
While providing maximum flexibility, this option requires changing both kernel and userspace code, and we limit our implementation to process-level domains for simplicity.

\section{\TheName Implementation}
\label{sec:impl}

Following \cref{sec:design}'s design principles, we implement the \TheName memory allocator in 1600 LoC and integrate it in Ubuntu 20 (kernel 5.15).
Our proof-of-concept implementation considers a typical server-grade memory system with 8KB DRAM rows, 128K rows per bank, and 128 banks spread across ranks and channels, resulting in 1MB per global row and 128GB of total memory capacity.
As discussed in \cref{sec:mapping}, a 4KB OS page is spread across multiple banks of the same global row ID, to improve bank-level parallelism. 
In other words, each \textit{global row}---spanning 128 memory banks---contains a set of 256 4KB OS pages. %
In our evaluation, we find that a chunk size of 16MB (i.e., 16 consecutive global rows) minimizes memory capacity overheads, striking a balance between memory loss and stranding.
Note that it is trivial to adjust the allocator's parameters to match other memory systems' characteristics.

\subsection{Memory Mapping}
\label{sec:mem_mapping_impl}

\usenix{
\TheName must track the mapping of frames (equivalently, logical rows) to physical rows in memory.
In simple DRAM mappings, a single frame is distributed across bank rows with the same row-ID~\cite{wang2020dramdig}. 
The row-ID bits are typically the highest-order bits of the physical address after accounting for columns, banks, ranks, and channels (e.g., bits 36 to 20), and are usually not hashed.
\TheName maintains a Global Row Table (GRT), a 256KB lookup table, that maintains the mapping of row-ID bit values that map to each global row location in memory.
In simple address mapping, this map is identity and a lookup table is not essential.
However, the GRT facilitates support for more complex DRAM addressing functions like scrambling, mirroring, and inversion, where row-ID bits \textit{are} hashed.
In such cases, a single frame can straddle several different physical DRAM rows. 
We elaborate on such mappings and their implications in \cref{sec:complex_addressing}. %
Inexpensive index-based GRT lookups are only required at memory allocation and de-allocation time.
}

\subsection{Kernel Interfaces}

\TheName is a memory allocator that replaces the buddy allocator.
We extend the memory initialization logic in the Linux kernel and override entry points into the buddy allocator, identified in six functions located in {\tt mm/page\_alloc.c} and {\tt mm/memblock.c}: 
{\tt free\_area\_init\-\_node()} and {\tt memblock\_\-free\_all()} used for memory initialization; 
{\tt \_\_rmqueue\_\-smallest()} and {\tt rmqueue()} used for page allocation; 
and {\tt \_\_free\_one\_page()} and {\tt free\-\_unref\-\_page\-\_commit()} used for page deallocation.
\revision{A new {\tt alloc\-\_from\-\_zonelet()} function enables explicitly requesting memory from a zonelet region.
Such functionality can be particularly useful in managing shared pages (currently disabled), as discussed in \cref{impl:sharing} and in Appendix~\ref{appendix:sharing}. 
}

\bgroup
\def\arraystretch{1.02}%
\begin {table}[tb]
\begin{scriptsize}
\begin{center} 
\caption{\revision{\TheName metadata overheads for a memory system with $C$ chunks, where $C=\frac{2^{M-N}}{X}$ for a system with $2^M$ byte capacity, $2^N$ byte global row, and X global rows per chunk.}}
\footnotesize

\begin{tabular}{|ll|p{1.3cm}|l|}
\hline
\multicolumn{2}{|c|}{\textbf{Metadata type}}                                   & \textbf{Mem. overhead (B)}       & \textbf{Comment}                               \\ \hline
\multicolumn{1}{|l|}{\multirow{4}{*}{\rotatebox{90}{Static}}}  & Chunk metadata            & \multicolumn{1}{|c|}{$C \times 544$} & Status flag, free page bitvec. \\ \cline{2-4} 
\multicolumn{1}{|l|}{}                         & Occupancy & \multicolumn{1}{|c|}{$C/8$}    & Chunk usage bitvec.                 \\ \cline{2-4} 
\multicolumn{1}{|l|}{}                         & GRT & \multicolumn{1}{|c|}{256K}    & Logical-to-physical row map                 \\ \cline{2-4} 
\multicolumn{1}{|l|}{}                         & Other                & \multicolumn{1}{|c|}{68}                             & Locks, domain list, stats             \\ \hline
\multicolumn{1}{|l|}{\multirow{3}{*}{\rotatebox{90}{Dynamic}}} & Per-domain           & \multicolumn{1}{|c|}{56}                             & Lock, usage flag, zonelist, stats    \\ \cline{2-4} 
\multicolumn{1}{|l|}{}                         & Per-zone             & \multicolumn{1}{|c|}{48}                             & Chunk list, stats                     \\ \cline{2-4} 
\multicolumn{1}{|l|}{}                         & Per-zonelet          & \multicolumn{1}{|c|}{0}                              & No \TheName-specific metadata                    \\ \hline
\end{tabular}

 \vspace{-4mm}
\label{table:metadata}
\end{center}
\end{scriptsize}
\end{table}
\egroup

\subsection{Bookkeeping}
\label{sec:impl:bookkeeping}

\TheName tracks active domains, the zones associated with each domain, the chunks comprising each zone, and the allocated/free pages within each chunk and zonelet.
For each zonelet and chunk, \TheName maintains a zonelet/chunk occupancy bitvector (using a bitmap in the kernel), where each bit represents the allocation status of a 4KB page. For 16MB chunks, a 512-byte bitvector tracks the chunk's 4096 contained pages. 
The bitvector is used to identify available space in each chunk and unused chunks that can be reclaimed.

\revision{\cref{table:metadata} summarizes \TheName's static and dynamic metadata, which incur negligible memory overhead. For a 128GB memory system with 1MB global row and a 16-row chunk, the static overhead is 4.26MB. The dynamic footprint depends on the workload's domains and resulting zones, but is at most a few MB even with thousands of domains and zones. }

\subsection{Shrinking and Splitting Zones}
\label{sec:design:opts}

\TheName attempts to create multi-chunk zones whenever possible to amortize memory capacity overheads (\cref{sec:design:zone_exp}).
While multi-chunk zones help reduce both memory loss and stranding for domains with growing memory footprint, they may exacerbate stranding when a domain is shrinking, because memory deallocations may result in large gaps of unallocated memory in a large zone.
\TheName leverages chunk occupancy bitvectors to resize or divide a multi-chunk zone as follows:
\begin{enumerate}[noitemsep,topsep=0pt,leftmargin=*]
    \item The entire first chunk {and} the second chunk's first $N_G$ global rows are unallocated. \TheName shrinks the zone by a chunk, by freeing the first chunk and converting the first $N_G$ global rows of the second chunk into guard rows.
    \item The zone's last chunk is unallocated. \TheName shrinks the zone by reclaiming the zone's last chunk.
    \item An entire chunk \textit{i} {and} the first $N_G$ global rows of chunk \textit{i+1} are unallocated. \TheName \textit{splits} the large zone into two smaller zones, by freeing chunk \textit{i} and converting the first $N_G$ global rows of chunk \textit{i+1} into guard rows.
\end{enumerate}

Zone manipulation actions involve only metadata updates without any data movement. 
While memory stranding can be aggressively tackled via compaction, it would not only increase complexity, but also introduce performance overheads of page migration (TLB shootdowns, page table updates, etc.). 
We found that our current implementation of opportunistic zone manipulation is sufficient to keep stranding low.

\subsection{Mapping Memory Allocations to Domains}

As described in \cref{sec:design:primitives}, \TheName's design supports high flexibility in the association between memory ranges and domains.
Our proof-of-concept implementation currently provides a simplified approach, where each Linux cgroup is implicitly associated with a different domain, and all memory allocation requests from a cgroup are associated with that cgroup's domain.
As explained in \cref{sec:design:zonelets}, a deployment-ready solution must at least isolate each page table page of the cgroup into a separate domain.
While we are in the process of implementing this isolation, we emulate additional domains for page tables in our prototype's evaluation, 
demonstrating \TheName's efficiency even for fine-grained domains (details in \cref{sec:method}).

\subsection{Memory Allocation: Zone or Zonelet?}

\revision{With the exception of {\tt alloc\_from\_zonelet()} invocations,}
upon each allocation request, \TheName decides whether to provide memory from a zone or a zonelet. Domains with very small memory footprint (KBs to couple of MBs) are best suited to zonelets, while zones are a better fit for larger domains. As the maximum size of a domain is not known a priori, \TheName uses a \textit{switch threshold} to switch between the two options: While a domain's memory footprint is below that threshold, all allocation requests are placed in zonelets. When that threshold is exceeded, new requests are placed in zones.
\revision{As a result, each domain initially only occupies space in zonelets, and as it grows beyond the switch threshold, its memory footprint straddles both zonelets and zones.}
Moreover, \TheName's metadata is exclusively placed in zonelets to disallow implicit attacks~\cite{zhang2020pthammer}, and we provide details on the bootstrapping process in Appendix~\ref{appendix:bootstrapping}.

\subsection{Memory Allocation/Deallocation Flow}

\begin{figure}[tb]
    \centering
     \includegraphics[width=.9\columnwidth]{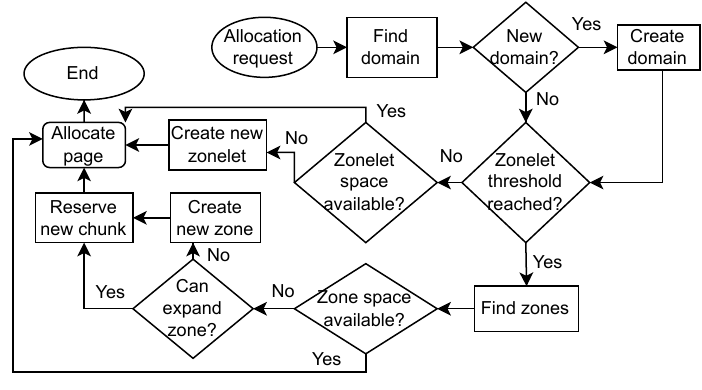}
    \caption{Flowchart of memory allocation logic.}    
    \label{fig:alloc}
\end{figure}

\begin{figure}[tb]
    \centering
     \includegraphics[width=.6\columnwidth]{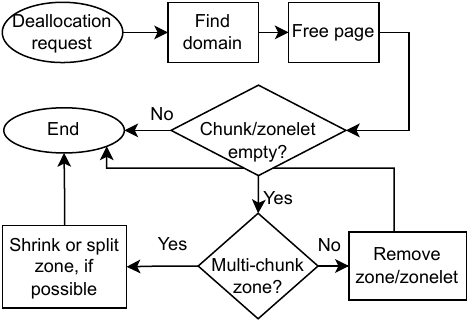}
    \caption{Flowchart of memory deallocation logic.}    
    \label{fig:dealloc}
\end{figure}

\cref{fig:alloc} shows the memory allocation flow. 
Depending on whether the domain's switch threshold has been met, \TheName searches for space in any available zonelets or in a chunk reserved for the target domain. \revision{{\tt{alloc\-\_from\-\_zonelet()}} requests skip the threshold check step and directly proceed to allocate memory from a zonelet.}
If no suitable zone or zonelet is found, a new chunk is reserved, and the required memory is allocated in the newly reserved space.
In that case, \TheName first attempts zone expansion before creating a new zone (\cref{sec:design:zone_exp}).
\usenix{Once the chunk the allocation will be serviced from is identified, \TheName can select a frame that maps in a specific global row within the chunk, by looking up the GRT. The GRT provides the row-ID bit values for frames that map at the desired global row.}
\cref{fig:dealloc} shows the deallocation flow. Page deallocation updates the respective occupancy bitvector, which may result in a zone shrinking/splitting opportunity. %

\section{Experimental Methodology}
\label{sec:method}

We evaluate \TheName against the baseline buddy allocator, which does not provide any isolation guarantees. We also evaluate recent DRAM isolation proposals---ZebRAM and Siloz---in a variety of workload scenarios to showcase their strengths and weaknesses. We conduct our experiments on a two-socket server with configuration parameters listed in \cref{table:system_config}. 
As discussed in \cref{sec:mapping,sec:impl}, we assume a standard memory mapping where each page is striped to different banks in memory onto rows sharing the same row ID.
With 128 banks and 8KB rows, each global row is 1MB in size.

\smallskip\noindent\textbf{Compared Configurations.}
We compare four memory allocators in terms of memory capacity overheads:
\begin{itemize}[noitemsep,topsep=0pt,leftmargin=*]
    \item The default \textit{buddy allocator}, %
    which serves as our baseline. 
    \item \textit{\TheName}, with $N_G=2$ and a 16MB chunk size comprising 16 global rows. %
    To zonelet-to-zone switch threshold is 12MB.
    \item \textit{ZebRAM}, approximated in \TheName by %
    exclusively using zonelets, each comprising two guard rows and one data row.
    We do not consider ZebRAM's optional use of guard space as swap storage with software integrity checks, as it incurs high performance overheads and is not secure (\cref{sec:prior-limitations}).
    \item \textit{Siloz}, which isolates domains by reserving entire DRAM sub-arrays to a domain. We  approximate Siloz with \TheName using $N_G=0$, chunk size = 512MB, and disabling zonelets.
\end{itemize}

\begin {table}[tb]
\begin{small}
\begin{center} 
\caption{System configuration.}
\footnotesize
\begin{tabular}{|c|c|}
\hline
   Processor           & 2 sockets, 24 logical cores/socket, 2.2GHz    \\ \hline
  Memory Size      & 128GB DDR4 at 2400MT/s \\ 
  Rows x Banks in Memory    & 128K $\times$ 128 \\
  Size of row and sub-array   & 8KB per row, 512 rows per sub-array \\ \hline
  OS / Kernel & Ubuntu 20.04.6 LTS / Linux 5.15.86 \\ \hline
\end{tabular}
  \vspace{-4mm}
\label{table:system_config}
\end{center}
\end{small}
\end{table}

\smallskip\noindent\textbf{Evaluation Metrics.}
We use memory capacity loss and memory stranding to compare the four configurations.
We do not explicitly compare in the evaluation is Rowhammer resilience, because the baseline buddy allocator is vulnerable to Rowhammer, while all three isolation schemes---\TheName, ZebRAM, and Siloz---provide the same guarantees by design.

\smallskip\noindent\textbf{Representative Workloads.}
We use SPEC2017 (int and fp)~\cite{SPEC2017} and Graph Analytics (GAP) \cite{GAP} workloads. SPEC workloads run in \textit{rate} mode with \textit{ref} dataset. GAP workloads use either the \textit{USA-road} graph or a synthetically generated kronecker graph. The resident memory footprint of SPEC workloads ranges from a few MBs to $\sim$1.2GB~\cite{singh2019memory} and we divide them into SPEC-small ($<$250MB), SPEC-medium ($<$750MB), and SPEC-large classes (SPEC-s/-m/-l). The footprint of GAP workloads is dictated by the input graph, which is 1.1GB for USA-road graph and 8GB for the synthetic graph.

\smallskip\noindent\textbf{Background Processes.}
We emulate background processes (kernel helpers, daemons, etc.), using synthetic workloads with a memory footprint that follows the distribution we observed across ten servers in our research cluster in both idle and busy systems. 
The servers feature 192--572GB of memory with an active memory footprint ranging from 1\% to 70\% of available memory. %
We observe 108 processes per server on average, of which 75 have a memory footprint $\leq$32MB. 
We mimic the observed background process memory footprint distribution %
by sampling an exponential distribution with an average $\mu=$4.9MB.
We also use this distribution to determine \TheName's switch threshold: set to 12MB, 90\% of small-footprint processes will be allocated exclusively in zonelets.

\smallskip\noindent\textbf{Page Tables.} 
Our current \TheName prototype creates one domain per \texttt{cgroup}. To capture the effect of separate domain creation for each page table (PT) page, each application (equivalently, cgroup)  with K MB of memory footprint additionally spawns K/2 dummy 4KB processes, each in a new domain/cgroup.
\revision{While doing so results in effectively double-counting the memory footprint of page tables, their magnitude relative to each application's data footprint is negligible.}

To ameliorate the limitation of its small number of supported domains, Siloz supports a special optimization whereby PTs are contained in their corresponding VM's domain but are protected by guard rows (similar to ZebRAM, or \TheName's zonelets).
To capture this effect in our Siloz emulation, we do not instantiate additional domains/cgroup for the PTs, but rather account for their memory footprint in terms of required memory usage and guard row loss.

\begin{table}[tb]
\begin{scriptsize}
\begin{center}
\caption{Workload mixes used in our experiments.} %

\begin{tabular}{|l|l|c|c|c|l|}
\hline
\multicolumn{1}{|c|}{\textbf{Mix}} & \multicolumn{1}{c|}{\textbf{Workloads}}                                                   & \multicolumn{1}{c|}{\textbf{\begin{tabular}[c]{@{}c@{}}Main\\ Apps\end{tabular}}} & \multicolumn{1}{c|}{\textbf{\begin{tabular}[c]{@{}c@{}}Backgrnd\\ Proc.\end{tabular}}} & \multicolumn{1}{c|}{\textbf{\begin{tabular}[c]{@{}c@{}}Page Table\\ Pages\end{tabular}}} & \multicolumn{1}{c|}{\textbf{Remarks}}                             \\ \hline
mix1                               & \begin{tabular}[c]{@{}l@{}}100\% \\ SPEC \& GAP\end{tabular}                              & 24                                                                                & 0                                                                                            & -                                                                                        & \begin{tabular}[c]{@{}l@{}}31GB mem.\\ 24 domains\end{tabular}     \\ \hline
mix2                               & \begin{tabular}[c]{@{}l@{}}80\% SPEC-s\\ 10\% SPEC-m \\ 10\% SPEC-l\end{tabular} & 128                                                                               & 0                                                                                            & -                                                                                        & \begin{tabular}[c]{@{}l@{}}32GB mem.\\ 128 domains\end{tabular}    \\ \hline
mix3                               & \begin{tabular}[c]{@{}l@{}}95\% SPEC-s\\ 5\% SPEC-m/l\end{tabular}    & 256                                                                               & 0                                                                                            & -                                                                                        & \begin{tabular}[c]{@{}l@{}}41GB mem.\\ 256 domains\end{tabular}    \\ \hline
mix4                               & \begin{tabular}[c]{@{}l@{}}50\% SPEC\\ 50\% GAP\end{tabular}                              & 24                                                                                & 0                                                                                            & -                                                                                        & \begin{tabular}[c]{@{}l@{}}65GB mem.\\ 24 domains\end{tabular}     \\ \hline
mix5                               & \begin{tabular}[c]{@{}l@{}}100\% \\ SPEC \& GAP\end{tabular}                              & 48                                                                                & 0                                                                                            & -                                                                                        & \begin{tabular}[c]{@{}l@{}}100GB mem.\\ 48 domains\end{tabular}    \\ \hline
mix6                               & \begin{tabular}[c]{@{}l@{}}90\% SPEC\\ 10\% GAP\end{tabular}                              & 48                                                                                & 512                                                                                          & $\sim$17,500                                                                             & \begin{tabular}[c]{@{}l@{}}35GB mem.\\ 18K domains\end{tabular}    \\ \hline
mix7                               & \begin{tabular}[c]{@{}l@{}}80\% SPEC\\ 20\% GAP\end{tabular}                              & 48                                                                                & 128                                                                                          & -                                                                                        & \begin{tabular}[c]{@{}l@{}}53GB mem.\\ 176 domains\end{tabular}    \\ \hline
mix8                               & \begin{tabular}[c]{@{}l@{}}100\% \\ SPEC \& GAP\end{tabular}                              & 48                                                                                & 128                                                                                          & $\sim$50,500                                                                             & \begin{tabular}[c]{@{}l@{}}101GB mem.\\ 51K domains\end{tabular} \\ \hline
mix9                               & \begin{tabular}[c]{@{}l@{}}100\% \\ SPEC \& GAP\end{tabular}                              & 48                                                                                & 512                                                                                          & $\sim$56,300                                                                             & \begin{tabular}[c]{@{}l@{}}112GB mem.\\ 57K domains\end{tabular} \\ \hline
mix10                              & 100\% SPEC                                                                                & 256                                                                               & 0                                                                                            & $\sim$52,800                                                                             & \begin{tabular}[c]{@{}l@{}}105GB mem.\\ 53K domains\end{tabular}   \\ \hline
\end{tabular}
 \vspace{-4mm}
\label{tab:mixes}
\end{center}
\end{scriptsize}
\end{table}

\smallskip\noindent\textbf{Workload Mixes.}
We generate ten mixes, each comprising a subset of the aforementioned workloads, background processes and page table processes. As shown in Table~\ref{tab:mixes}, the mixes comprise real-world scenarios to test different aspects of the domain-isolation schemes, like scaling to thousands of domains as well as low and high memory usage.
Each process runs in a separate \texttt{cgroup}.
\revision{We run each workload mix for two hours. Each workload mix restarts execution upon completion, and as workload runtimes vary, our experiments capture highly dynamic execution scenarios. To illustrate, each of mix1's workloads executes an average, maximum, and minimum of 88, 382, and 10 times, respectively.}

\section{Evaluation}
\label{sec:eval}
\begin{figure*}
    \centering    
     
        \includegraphics[width=.9\textwidth]{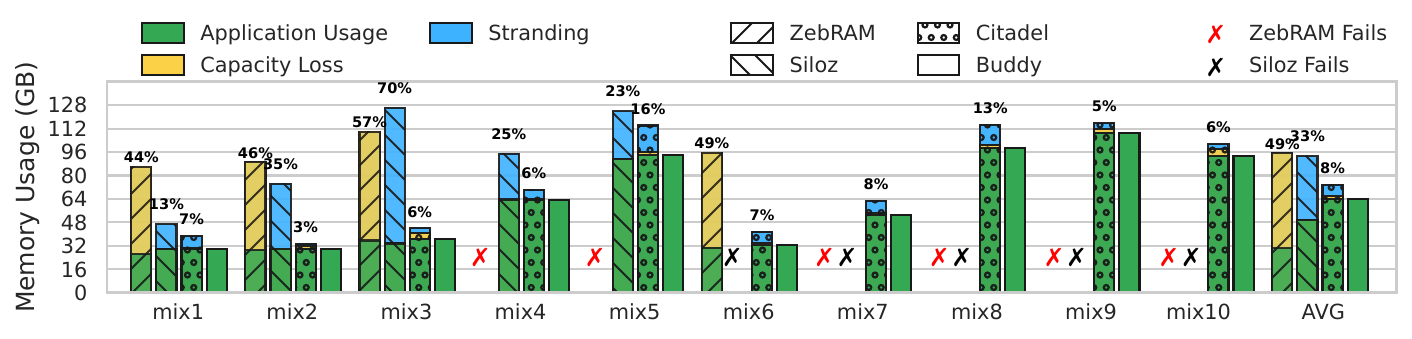}
    \caption{Breakdown of memory usage of ZebRAM, Siloz, and \TheName, and Buddy allocator into application usage, guard row capacity loss, and stranded memory. Over-bar numbers denote memory overhead (stranding + loss) compared to Buddy as a fraction of 128GB memory capacity. Prior works impose drastic overheads and fail for several mixes. \TheName works for all mixes and retains lowest overheads. \textit{Note that ``Avg.'' for ZebRAM and Siloz only includes the small-footprint mixes they support.}}
    \label{fig:main-eval}
\end{figure*}

\begin{figure*}
    \centering    

         \includegraphics[width=\linewidth]{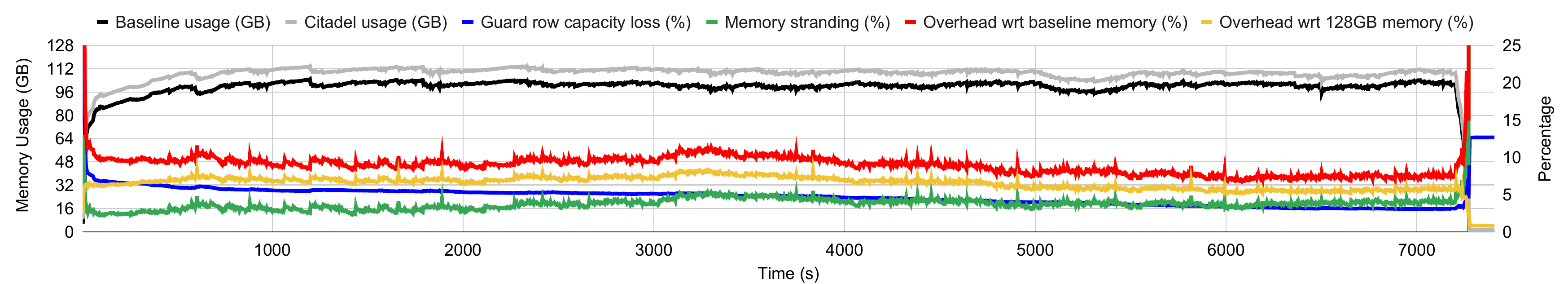} 
    \caption{Execution timeline for workload mix10 showing absolute memory utilization (primary y axis), as well as relative memory capacity loss and stranding (secondary y axis). }
    \label{fig:timeline}
    
\end{figure*}

\subsection{Memory Stranding and Waste}
\label{sec:eval:main}

\cref{fig:main-eval} summarizes the memory overheads introduced by \TheName as well as Siloz, ZebRAM, compared to the baseline Buddy allocator.
Each stacked bar breaks down the total memory usage into application usage, memory loss due to guard rows, and memory stranding. The Buddy allocator's total memory usage is equivalent to application usage. 

\TheName not only reduces memory overheads drastically compared to ZebRAM and Siloz, but is also the only Rowhammer-aware allocator that supports all 10 workload mixes.
ZebRAM only supports 4/10 workloads due to excess memory loss, while Siloz supports 5/10 due to excess stranding.
Considering only the mixes supported by all allocators, \TheName yields, on average, a {$9.8\times$} and {$4.6\times$} lower overhead than ZebRAM and Siloz, respectively.
\TheName introduces a modest memory overhead of \aegisAvgOverhead on average, and up to 15.8\%.
Except for mix 3 and 10, the overhead is virtually exclusively attributed to stranding rather than loss. 
\revision{In all cases, the overheads attributed to \TheName metadata are negligible, as they account for at most a couple of MBs (cf. \cref{sec:impl:bookkeeping}).}

\subsection{Performance}
\label{sec:eval:perf}

For each of \cref{sec:eval:main}'s experiments, the time spent in \TheName and buddy memory allocator logic accounts for $<1\%$ of the experiment's runtime, and is therefore insignificant. 
We also do not observe contention due to locking and busy-waiting as we employ fine-grained per-domain locks.
Hence, \TheName provides domain-isolation without any performance loss.

\subsection{Temporal Behavior}
\label{sec:eval:temporal}

\cref{fig:timeline} shows mix 10's temporal execution behavior over two hours.
The black and gray lines indicate the difference between the memory requested by the workloads versus the memory usage \TheName results in.
To analyze the memory usage gap, \cref{fig:timeline} also shows \TheName's incurred memory overhead relative to the workload mix's requested memory and to the system's total memory (128GB). The memory overhead is further broken down into stranding and loss.

The memory overhead corresponds to about 6\% of the total system capacity, or 9\% over the memory required by the workload mix. The overhead is roughly equally split between stranding and loss, indicating that a chunk size of 16 rows offers a good balance between the two overhead sources.
The overhead remains relatively constant throughout the execution, except ramp-up and ramp-down, where the \textit{relative} overhead briefly skyrockets as high as  60\%. However, their duration is brief, and their \textit{absolute} represents a negligible fraction of the total memory capacity, as indicated by the orange line.

\subsection{Sensitivity Studies}
\label{sec:eval:sensitivity}

\subsubsection{Impact of \TheName Optimizations}

\begin{figure}
    \centering    
\captionsetup[subfloat]{captionskip=-2pt}

  \subfloat[Incremental effect of \TheName optimizations.] {
                \includegraphics[width=.95\columnwidth]{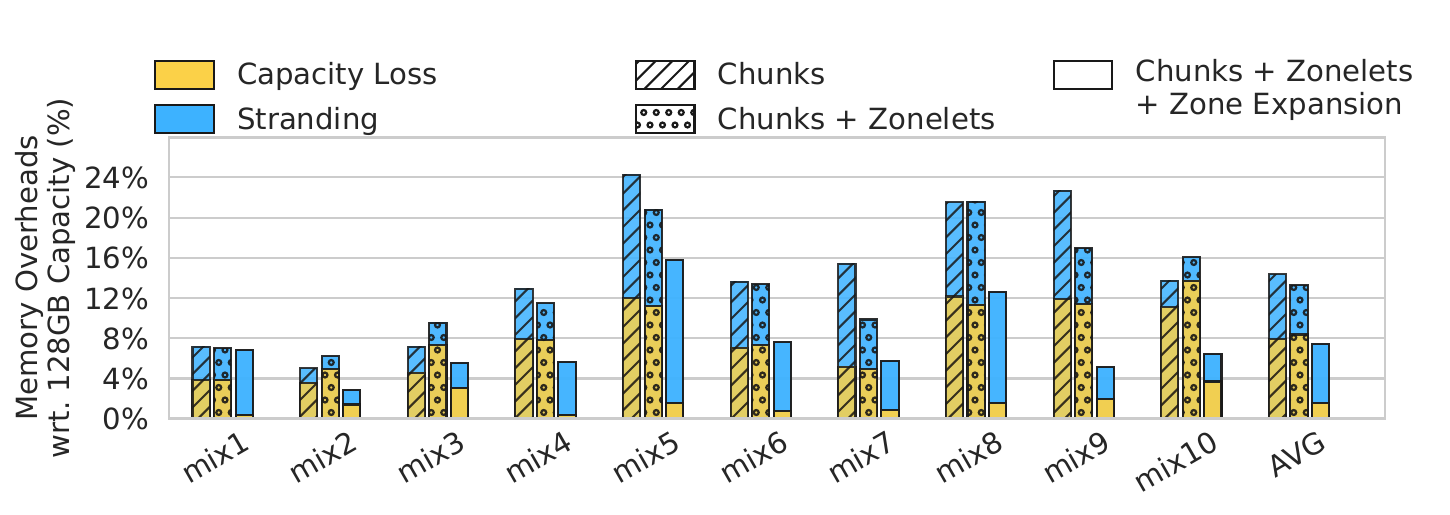}
    		\label{fig:aegis-opts}
    	}
           
     \subfloat[Effect of chunk size on memory loss and stranding.]{ %

              \includegraphics[width=.95\columnwidth]{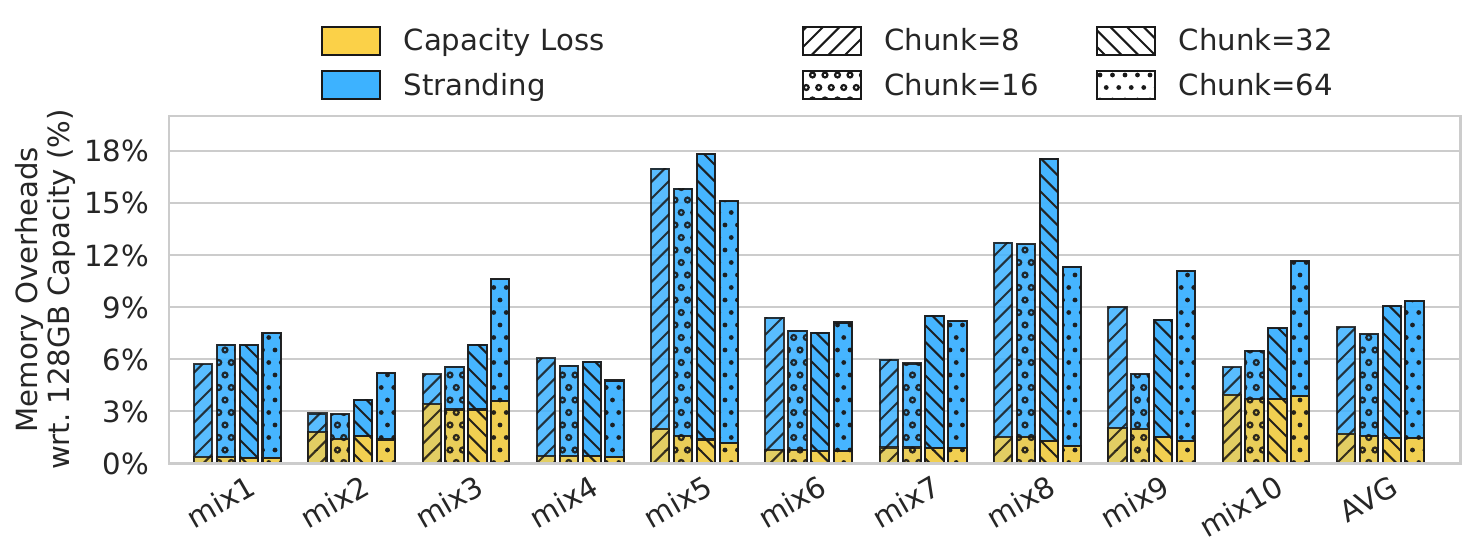}
    	    \label{fig:chunk-size}                
    	}

        \subfloat[Effect of guard row count ($N_G$) on memory loss and stranding.]{ %
                \includegraphics[width=.95\columnwidth]{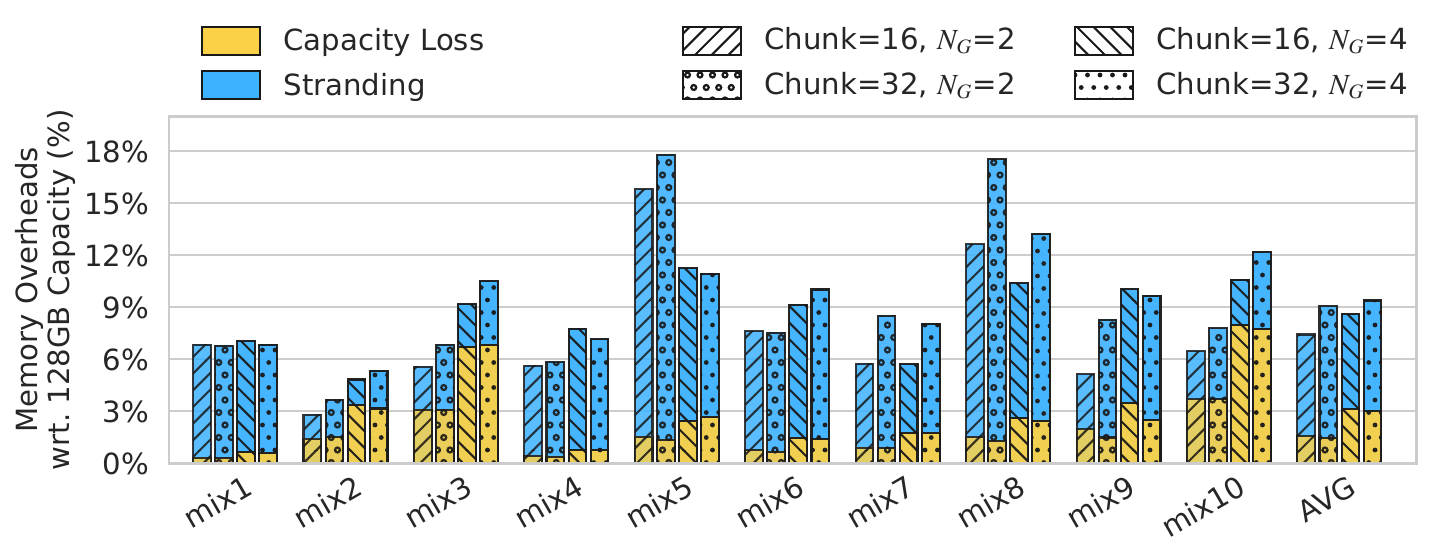}
    		\label{fig:guard-row}
    	}
    \caption{Sensitivity studies. }
    \label{fig:sensitivity}
\end{figure}

\cref{fig:aegis-opts} breaks down \TheName's optimizations. \TheName using only 16-row chunks offers memory overhead of about 14\% on average, split evenly between capacity loss and stranding.
Optimizations outlined in \cref{sec:design:zone_exp} and \cref{sec:design:zonelets} further reduce these overheads.
Zonelets primarily protect small, programmer-defined bespoke domains, which our evaluations lack.
However, even the relatively larger domains belonging to background processes (with an average memory footprint of 5MB) benefit from zonelets, reducing overhead to 13\%.
Without zone expansion, the capacity loss due to guard rows increases significantly,
accounting for more than half of the 13\% average overhead being attributed to guard row loss.
Zone expansion lowers capacity loss to just 2\% of the DRAM capacity, and total memory overheads to 7\% on average.
Thus, each optimization contributes toward lowering the memory overheads.

\subsubsection{Impact of Chunk Size}

\cref{fig:chunk-size} provides \TheName's memory loss and stranding overheads (compared to 128GB capacity) at varying chunk sizes.
As chunk size increases, the overhead due to capacity loss decreases slightly, but is remarkably low (2\% on average) even for chunk size of 8, which would have worst-case loss of 25\% ($N_G=2$).
We attribute low capacity loss to \TheName's zone expansion which opportunistically creates large, contiguous zones.
Stranding broadly increases with chunk size, as the likelihood of unallocated memory in a reserved chunk increases with a larger chunk, with a minima at chunk size of 16.
\TheName's default chunk size is 16, offering the lowest overheads while supporting up-to 8K large domains.

\subsubsection{Impact of Guard Rows}

\TheName assumes $N_G=2$ guard rows are sufficient to prevent Rowhammer bit flips even with complex attack patterns, in line with prior works~\cite{half-double}.
However, DRAM's worsening vulnerability to Rowhammer might necessitate more guard rows, with $N_G=4$ being recommended in a recent study~\cite{lang2023blaster}.
We showcase \TheName's adaptability to such trends in \cref{fig:guard-row}, which plots overheads of \TheName with chunk sizes \{16, 32\} and $N_G=\{2,4\}$.
\revision{Given a chunk size, increasing $N_G$ usually leads to higher overheads: while capacity loss always increases, stranding may reduce in some cases (e.g., mix5,7,8). Higher $N_G$ results in fewer data rows per chunk. In some workloads, this effect aids chunk freeing/splitting upon dynamic deallocation, reducing stranding.}
On average, with $N_G=4$, \TheName's overhead increases but remains low at %
9\% of DRAM capacity.
At $N_G=4$, larger chunk size of 32 performs marginally worse. %
Therefore, \TheName scales gracefully to future systems, maintaining low overheads.

\section{Discussion}
\label{sec:discussion}

\subsection{Internal DRAM Addressing Implications}
\label{sec:complex_addressing}

\begin{figure}
    \centering    
           \includegraphics[width=.95\columnwidth]{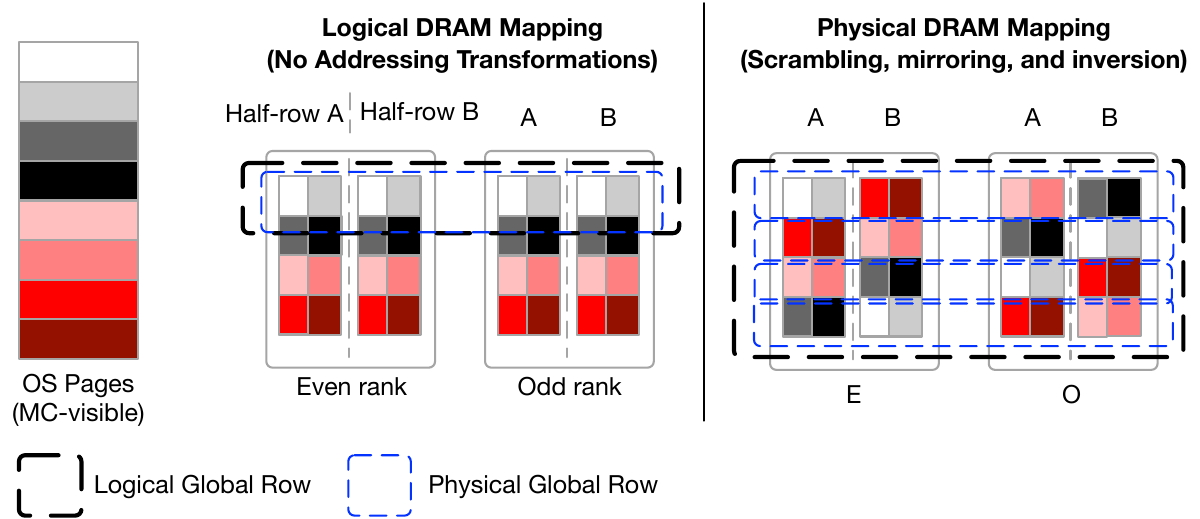}
    \caption{\revision{Row address transformations to logical DRAM mapping distributes the same row (and, consequently, portions of OS frame) to four distinct physical rows in two ranks.}}
    \label{fig:complex_addressing}
\end{figure}

\usenix{
For simplicity, we explained key tenets of \TheName assuming the DRAM rows are not scrambled, mirrored, or otherwise re-arranged internally.
\TheName is able to tackle these complexities that are common in modern servers.
We defined global row as the set of DRAM rows that share the same row-ID bits. 
In absence of internal DRAM addressing, the same row-ID maps to the same physical row across all memory banks.
With row mirroring and inversion, the same global row can reside in up to four physical locations, depending on the rank and DRAM row side (each row is managed as two half-rows)~\cite{loughlin2023siloz}. 
Notably, mirroring and inversion do not impact the three least significant bits of row-ID, so groups of 8 rows retain their relative position. 
However, row scrambling hashes the lower three significant bits of the row-ID, changing the relative positions of DRAM rows \textit{within a group of 8 DRAM rows}.
\cref{fig:complex_addressing} illustrates the impact of these transformations with an example. Each color represents a different OS page.

Analysis of these transformations, detailed in Appendix~\ref{appendix:dram_addressing}, reveals \textit{two key invariants}. First, even though a global row is spread into four physical locations, the set of row-IDs that map to these physical locations is the same. That is, a set of four global rows is rearranged to occupy portions of four physical DRAM row locations. \TheName manages each group of four related global rows (of 1MB each) as a single \textit{logical global row}. Second, as scrambling and inversion/mirroring impact different bits in row-ID, the row ordering within a group of 8 DRAM rows remains the same, even when they are spread to multiple physical locations. That is, the $i_{th}$ physical row in each of a logical global row's four physical locations will be occupied by frames with the same row-ID values (albeit at a different order in the horizontal dimension).
Therefore, \TheName's concept of chunk, which is a collection of contiguous (logical) global rows, as well as guard rows being the first $N_G$ rows within a chunk, remains well-defined, although the chunk is physically spread to four locations.

Thus, \TheName's tenets of operation remain intact, with few logistical changes.
The memory system (of 128K global rows) now comprises 32K logical global rows. Moreover, as contiguity beyond groups of 8 logical global rows is not guaranteed, it sets a natural upper bound for the chunk size (32MB). Finally, with $N_G = 2$, the guard row overhead increases to 25\% of the 8 logical global rows comprising the chunk. The Global Row Table (GRT) facilitates mapping of global row-IDs that comprise each logical global row, and only requires four 2-Byte physical global row-ID mappings for each of the 32K logical global rows, totaling 256KB. The GRT is looked up during allocation to find the PFN, given the relative page-index from the chunk occupancy bitvector. Moreover, during de-allocation, to find the chunk for the to-be-freed PFN, either the GRT can be searched (as de-allocation is lazy), or a 256KB inverse-GRT can be provisioned. %

The implications of complex addressing are two-fold. The spread of chunks to multiple physical locations means that a chunk can have more than two neighbors. We find that, for the transformations outlined in prior works~\cite{cojocar2021mfit, loughlin2023siloz}, 33\% of the chunks have two neighbors, 33\% have three neighbors, and 33\% have four neighbors. \TheName's zone expansion optimization remains applicable, but expanding a chunk with N neighbors requires finding N-1 of them free, reducing chances of success. 
As a result, with additional restrictions on zone expansion, capacity loss on guard rows becomes more likely to dominate and approach memory overheads of 25\%, compared to our demonstrated 7.4\% in absence of complex addressing.

}

\subsection{\revision{Inter-Process Memory Sharing}}
\label{impl:sharing}

Memory sharing reduces memory pressure and aids inter-process communication. However, shared memory can lead to cross-domain memory accesses. Therefore, it is commonly disabled to eliminate many attack vectors (like cache side channels~\cite{yarom2014flush+}), including Rowhammer.
In our experiments, memory sharing opportunities are low: $<$0.1\% of the OS pages have a \textit{refcount} $>$1 on average (maximum of 0.41\%, details in Appendix~\ref{appendix:sharing}). 
As a result, we disable memory sharing in \TheName by default to simplify our implementation, with negligible overheads, arising chiefly due to duplication of previously shared libraries.
Nonetheless, the OS can be extended to support Rowhammer-resilient shared memory at low cost with \TheName's support, by placing all shared pages in zonelets.
Such extension requires modest kernel modifications beyond the memory allocator level.
While not implemented in our current prototype, we outline its design in Appendix \ref{appendix:sharing}.

\section{Related Work}
\label{sec:related}

\smallskip
\noindent\textbf{Rowhammer protection in hardware.}
\label{sec:related:hardware}
Low-cost hardware defenses have been developed by industry~\cite{samsung_dsac, isscc23}. Unfortunately, they trade off security for area efficiency and cannot track all aggressors, leading to attacks even in presence of commercially deployed mitigations like TRR~\cite{frigo2020trrespass}.
Principled hardware schemes~\cite{park_graphene:_2020, qureshi2022hydra, wi2023shadow} can track all aggressors. However, \revision{hand-crafted patterns like Half-Double~\cite{half-double}} and worsening DRAM reliability can render proposed mitigations~\cite{saileshwar2022RRS} vulnerable to unforeseen breakthrough attacks~\cite{cojocar2019eccploit, srs, ali2022safeguard}. Moreover, such mitigations are yet to be deployed~\cite{loughlin2023siloz}, leaving current and future devices vulnerable. 
\usenix{Recently, JEDEC revealed a framework to support Per-Row Activation Counters (PRAC)~\cite{JEDEC-prac}. 
Unfortunately, PRAC adds 9\% area overhead in DRAM~\cite{samsung_dsac} and makes every DRAM activation a read-modify-write operation, incurring significant performance loss. 
PRAC is optional for DRAM vendors to implement and, given its limitations, widespread adoption remains unlikely.}

\smallskip
\noindent\textbf{Rowhammer protection in software.}
\label{sec:related:software}
ANVIL~\cite{aweke2016anvil} tracks aggressor row accesses using performance counters, but it is vulnerable to Half-Double~\cite{HalfDouble} attack. 
GuardION~\cite{van2018guardion} uses one guard row between data of different security domains, but suffers from drastic performance loss similar to ZebRAM.
\revision{RIP-RH~\cite{bock2019rip} provides Rowhammer-aware memory allocation, but  places the 4KB page in one DRAM row and would be incompatible with typical mappings that distribute a page across memory banks. Such mappings necessitate the global-row concept.}
CATT~\cite{brasser2017can} blacklists pages with vulnerable DRAM cells, but in modern memory, more than 90\% of the caapcity would be unavailable with this approach~\cite{monotonic}.

\smallskip
\noindent\textbf{Victim placement in Rowhammer attacks.}
Massaging victim pages into vulnerable rows adjacent to attacker controlled rows is critical for success of Rowhammer exploits~\cite{frodo, pessl2016drama, xiao2016bitflipflops}. 
Examples of memory massing include inducing victim to spawn numerous processes~\cite{seaborn2015exploiting}, and exploiting policies like page deduplication~\cite{flipfengshui}, per-CPU page frame cache \cite{tobah2022spechammer}, or the page allocation itself~\cite{van2016drammer}.
By placing each exploited entity in a separate domain, \TheName can protect against all such attacks by preventing their root-cause, which is the adjacency of untrusting entities (domains) in memory.

\section{Conclusion}
\label{sec:conclusion}

We present \TheName, \revision{a memory allocator capable of isolating an arbitrary number of software-defined security domains of any size.}
\TheName incurs no performance penalty and, as a software-only solution, is deployable on most systems. 
Leveraging the primitives of reservation chunks, zones, and zonelets, \TheName effectively tackles the two sources of memory capacity overhead---loss and stranding---for security domains sized from KBs to GBs, while
preserving isolation guarantees for up to millions of domains.
We demonstrated that \TheName enables workload deployments that are not supported by prior software-based solutions, and guarantees Rowhammer resilience at modest memory capacity overheads of just 7.4\% on average across SPEC and GAP workloads.

\newpage

\bibliographystyle{plain}
\bibliography{references}
\appendix

\section{\TheName Glossary}

\cref{table:glossary} summarizes the key terms used in \TheName's design.

\def\arraystretch{1.02}%
\begin {table}[h]
\begin{scriptsize}
\begin{center} 
\caption{Key terms used in \TheName.}
\footnotesize
 
\begin{tabular}{|p{13mm}|p{5.5cm}|}
\hline
\textbf{Term}    & \textbf{Description}     \\ \hline
Security domain & A user-defined memory footprint that must be protected from Rowhammer attacks by entities outside that domain. In our current implementation, we assume a one-to-one correspondence of a process to a security domain, but finer-grained domain definition is possible.\\ \hline
Security zone & A contiguous memory region belonging to the same domain. A domain may comprise multiple zones.\\ \hline
Global row & The set of DRAM rows sharing the same row ID across all banks in a NUMA domain’s memory.\\ \hline
Guard row & A global row reserved for security purposes, not accessible by any user process.\\ \hline
Reservation chunk & \TheName's unit of memory reservation. A chunk comprises a configurable number of contiguous global rows.\\ \hline
Logical global row & A set of global rows that get rearranged to occupy portions of the same physical row locations.\\ \hline
Global Row Table (GRT) & A lookup structure that records all the address row-ID values that map to the same logical global row. \\ \hline
\end{tabular}

\label{table:glossary}
\end{center}
\end{scriptsize}
\end{table}

\section{DRAM Addressing Transformations}
\label{appendix:dram_addressing}
We refer to row-IDs supplied by the memory controller as ``logical'' row addresses and those used by the DRAM to physically access rows as ``physical'' row addresses. 
Aside from remapping of faulty rows, prior works~\cite{loughlin2023siloz, cojocar2021mfit} have identified row address scrambling, mirroring, and inversion as key logical-to-physical address transformations within a bank.

\smallskip
\noindent\textbf{Row-address Scrambling.} Similar to bank hashing, address scrambling derives the physical row address by XOR-ing several logical row-address bits. For example, Cojocar et al.~\cite{cojocar2021mfit} found that groups of 8 logically contiguous rows are scrambled in some DRAM modules, with higher-order bits remaining unchanged. While scrambling changes the relative order of rows within a group of 8 rows, their overall contiguity is maintained. Moreover, the same scrambling function is used across all banks in memory. Overall, despite scrambling, the position of each row can be identified, thus isolation primitives like guard rows remain viable.

\smallskip
\noindent\textbf{Row-address Mirroring and Inversion.} Siloz~\cite{loughlin2023siloz} provides detailed description of these transformations, which we explain briefly for brevity. Row-address mirroring swaps several pairs of logical bits in \textit{odd-numbered} ranks within a DIMM to get the physical row-ID. Meanwhile, address inversion inverts several bits in one side of half-rows of ranks in the DIMM. Notably, mirroring and scrambling do not impact the three least significant bits of row-address. Thus, the same logical row, across a pair of odd and even ranks in the DIMM, might reside in up to four different physical half-row locations. Row inversion and mirroring are commutative transformations, so while row-IDs are distributed to up-to four physical locations, each of these locations contain the same four row-IDs.  

\section{Secure Memory Sharing}
\label{appendix:sharing}

To enable secure memory sharing, we leverage our observation that co-located pages within the same row that are flanked by guard rows cannot hammer each other.
Therefore, secure sharing can be achieved by placing all pages that are shared in the zonelet region of \TheName, which requires changes to OS' page fault mechanism.
By default, the OS does not invoke the memory allocator if a process attempts to \texttt{mmap} a page that is already allocated to another process, and instead updates the requesting process' page table to point to the allocated PFN.

To support secure sharing, this flow would be minimally modified, wherein on a minor page fault for a to-be-shared page, the OS explicitly requests a new frame from a zonelet region.
It then performs a typical page migration to the new frame, as is done in multi-tier or NUMA systems, which involves a \texttt{memcpy}, TLB shootdown, and page table updates for affected processes.
Subsequent requests to map an already shared page are served via the default fast path, as the page was previously moved to a zonelet.

\begin{table}[htb]
\begin{scriptsize}
\begin{center}
\caption{\revision{Pages moved as fraction of allocated pages of the mix and associated migration rate of secure memory sharing.}} %
\revision{
\begin{tabular}{|l|l|l||l|l|l|}
\hline
\textbf{Mix} & \multicolumn{1}{c|}{\textbf{\begin{tabular}[c]{@{}c@{}}\% Pages \\ Moved\end{tabular}}} & \multicolumn{1}{c||}{\textbf{\begin{tabular}[c]{@{}c@{}}Migration\\ Rate (KB/s)\end{tabular}}} & \textbf{Mix} & \multicolumn{1}{c|}{\textbf{\begin{tabular}[c]{@{}c@{}}\% Pages\\ Moved\end{tabular}}} & \textbf{\begin{tabular}[c]{@{}l@{}}Migration\\ Rate (KB/s)\end{tabular}} \\ \hline
mix1         & \textless{}0.01\%                                                                       & 8.5                                                                                           & mix6         & \textless{}0.01\%                                                                      & 0.2                                                                      \\ \hline
mix2         & 0.23\%                                                                                  & 213.1                                                                                         & mix7         & 0.01\%                                                                                 & 41.3                                                                     \\ \hline
mix3         & 0.41\%                                                                                  & 0.3                                                                                           & mix8         & 0.02\%                                                                                 & 143.8                                                                    \\ \hline
mix4         & \textless{}0.01\%                                                                       & 312.5                                                                                         & mix9         & 0.02\%                                                                                 & 234.5                                                                    \\ \hline
mix5         & 0.04\%                                                                                  & 0.2                                                                                           & mix10        & \textless{}0.01\%                                                                      & 0.8                                                                      \\ \hline
\end{tabular}
}
\label{table:sharing}
\end{center}
\end{scriptsize}
\end{table}

\medskip
\noindent\textbf{Impact of Inter-process Memory Sharing.}
To evaluate the degree of memory sharing, we measure when a page's \texttt{refcount} is increased from one (which means page is private) over the two-hour duration of evaluated workloads.
\cref{table:sharing} shows that to enable secure memory sharing, on average, just 0.07\% of the pages would need to be moved by \TheName to a zonelet, which translates to an average data-movement rate of just 117KB/s. %
Moreover, the performance impact due to ensuing TLB shootdown would be negligible.

\section{\TheName Bootstrapping}
\label{appendix:bootstrapping}

To avoid advanced attacks that can implicitly target critical data~\cite{zhang2020pthammer}, \TheName's data structures always reside in zonelets, thereby disallowing Rowhammer exploits.
\TheName, along with a small part of the Linux Kernel, resides in contiguous memory for a portion of system startup.
During boot up, the Linux kernel loads the buddy allocator and its data structures as part of the \texttt{free\_area\_init\_core} function. 
In the \texttt{memblock\_free\_all} function, where buddy's free list structures are built, we initialize \TheName's data structures.
Once they are initialized---and before any userspace processes are initiated---the kernel moves them to zonelets. 
Subsequent \TheName metadata is provisioned exclusively via zonelets.

Once \TheName is bootstrapped, the kernel segments occupy multiple zones (with requisite guard rows) within the single kernel domain, protecting it from subsequent memory allocation requests from userspace processes. 
Provisioning one domain per kernel thread is a straightforward extension.
Moreover, similar to bootstrapping \TheName, the entire kernel (with limited exceptions, such as immovable memory regions) can be relocated to zonelets for complete protection.

\end{document}